# Magnetic and magnetocaloric effect in a stuffed honeycomb polycrystalline antiferromagnet GdInO$_3$

Yao-Dong Wu(吴耀东)[a,b,c], Wei-Wei Duan(段薇薇)[a], Qiu-Yue Li(李秋月)[a], Yong-Liang Qin(秦永亮)[b], Zhen-Fa Zi(訾振发)[a,*], Jin Tang(汤进)[b,*]

[a] *Universities Joint Key Laboratory of Photoelectric Detection Science and Technology in Anhui Province, Anhui Province Key Laboratory of simulation and design for Electronic information system, and School of Physics and Materials Engineering, Hefei Normal University, Hefei, 230601, China*

[b] *Anhui Province Key Laboratory of Condensed Matter Physics at Extreme Conditions, High Magnetic Field Laboratory, HFIPS, Anhui, Chinese Academy of Sciences, Hefei, 230031, China*

[c] *University of Science and Technology of China, Hefei, 230026, China*

* Corresponding Author.

E-mail Address: zfzi@issp.ac.cn (Z. F. Zi); tangjin@hmfl.ac.cn (J. T.)

**Abstract**

The magnetic and magnetocaloric properties were studied in a stuffed honeycomb polycrystalline antiferromagnet $GdInO_3$. The onset temperature of antiferromagnetic ordering was observed at ~ 2.1 K. Negligible thermal and magnetic hysteresis suggest a reversible magnetocaloric effect (MCE) in the $GdInO_3$ compound. In the magnetic field changes of 0–50 kOe and 0–70 kOe, the maximum magnetic entropy change values are 9.65 J/kg K and 18.37 J/kg K, respectively, near the liquid helium temperature, with the corresponding relative cooling power values of 115.01 J/kg and 211.31 J/kg. The MCE investigation of the polycrystalline $GdInO_3$ serves to illuminate more exotic properties in this frustrated stuffed honeycomb magnetic system.

# Introduction

With the increasingly severe issues of global warming and the energy crisis, the magnetic refrigeration (MR) technique based on the magnetocaloric effect (MCE) of magnetic materials has revealed superior advantages such as higher energy efficiency, a more robust design, and greater environmental friendliness compared to the conventional gas-compression refrigerant methods using ozone-consuming volatile refrigerants.[1-3] The search for high-performance magnetocaloric materials has significant meaning in the future applications of the MR technique at room temperature for domestic and industrial purposes,[4,5] with the representative compounds such as Gd metal,[6] La(Fe, Si),[5] $Gd_5(SiGe)_4$,[7] MnFe(P, As),[8] FeRh,[9] and La$A$MnO$_3$ ($A$ = $Ca^{2+}$, $Ba^{2+}$, $Na^+$, $K^+$, and so on).[10-13] In addition, it is important to explore magnetocaloric materials at the cryogenic temperature, which would be beneficial for hydrogen or helium liquefaction and space science.[5,14] Over the past few decades, attention has mostly been paied to cryogenic magnetocaloric materials with rare-earth-based alloys and oxides, such as $Gd_3B_5O_{12}$ ($B$ = Ga, Fe, Al) [15], $RM_2$ ($R$= rare earth elements, $M$ = Al, Ni, Co),[16-18] $RM$ ($M$ = Zn, Ga),[19-22] $RMX$ ($M$ = Fe, Co, X = Al, Mg, C),[23,24] $R_2T_2X$ ($T$ = Cu, Ni, Co, X = In, Al, Ga, Sn, and so on),[25] $R_{60}Co_{20}Ni_{20}$ ($R$ = Ho and Er),[26] $La_{1-x}Pr_xFe_{12}B_6$,[27] $Gd_{20}Ho_{20}Tm_{20}Cu_{20}Ni_{20}$,[28] DyNiGa,[2] dual-phase HoNi/HoNi$_2$ composite,[29] $RNO_3$ ($N$ = Al, Fe, Mn, Cr, and so on),[30-37] $R_2M_2O_7$.[38-42] In particular, recent studies have demonstrated that the $Gd^{3+}$ and $Eu^{2+}$ ion-based compounds display great MCE performances due to the large angular momentum of the half-filled 4$f$ shell (4$f^{\,7}$) and negligible crystal electrical field (CEF) effect with $J = S = \frac{7}{2}, L = 0$, with the representative compounds such as GdFeO$_3$,[43] GdScO$_3$,[44] GdCrO$_3$,[45] GdAlO$_3$,[46] GdVO$_4$,[47,48] GdPO$_4$,[49] GdBO$_3$,[50] EuTiO$_3$.[51,52]

In contrast to conventional magnetic systems with long-range ferromagnetic (FM) or antiferromagnetic (AFM) ordering, geometrically frustrated magnets possess peculiar lattice structures, leading to a competition between the neighboring spin-spin interactions. Exotic magnetic properties such as spin ice and the spin liquid state can be expected in geometrically frustrated magnets with Kagome, garnet, and pyrochlore lattices.[53] As a result, disordered cooperative paramagnetic ground states can remain even at temperatures much lower than the paramagnetic Curie-Weiss temperature $\theta_w$. Many geometrically frustrated magnets have been reported to possess considerable magnetocaloric performances in cryogenic temperature regions, since the quasi-paramagnetic ground states in these systems contribute no reduced entropy, which can hinder the temperature change in the adiabatic demagnetization process.[53] The typical cases are antiferromagnet $R_2M_2O_7$ ($M$ = Ti, Mo) with pyrochlore lattices,[39,40,42,54] $Gd_3Ga_5O_{12}$ with garnet lattice,[55] $R_3BWO_9$ with distorted Kagome magnetic lattices,[56,57] and $SrGd_2O_4$ with distorted honeycomb magnetic lattices,[58] as well as $TmMgGaO_4$,[59] $RBO_3$,[50] and $Ba_3Ln(BO_3)_3$ ($Ln$ = Ho–Lu) with two-dimensional (2D) triangular magnetic lattices.[60]

Rare-earth indium oxides $R$InO$_3$ possess three main types of crystal structures.[61] As the radius of rare-earth ions decreases, the crystal structures can be orthorhombic with the space group *Pnma* (for $R$ = La, Pr, Nd, and Sm), hexagonal with the space group *P6$_3$cm* (for $R$ = Sm-Ho, Y), or cubic with the space group $Ia\overline{3}$ (for $R$ =Ho, Er, Yb). Hexagonal $R$InO$_3$ displays fascinating properties such as pressure-induced hexagonal to orthorhombic crystal structure transformation,[62] geometric multiferroic behavior,[63] and negative thermal expansion.[64] Notably, the crystal structure of hexagonal $R$InO$_3$ consists of 2D triangular layers of $R^{3+}$ ions separated by non-magnetic layers of corner-sharing [InO$_5$] trigonal bipyramids,[65] as shown in Fig. 1(a). This unique magnetic lattice, also called the stuffed honeycomb lattice, is formed by two nonequivalent rare-earth $R^{3+}$ sites with the $R$1 ions in the honeycomb sites and $R$2 at the center of each hexagon (see Fig.1(b)).[66] If anisotropic antiferromagnetic interactions caused by spin-orbit coupling exist between the two $R^{3+}$ sites, also known as bond-dependent Kitaev interactions, a frustrated spin liquid ground state without long-range magnetic ordering can be expected in the honeycomb magnetic system.[67,68] As one crucial member of hexagonal $R$InO$_3$, TbInO$_3$ has recently been revealed to be a highly frustrating spin liquid magnetic system with a stuffed $Tb^{3+}$ ion-based honeycomb lattice,[67,69] which is very extraordinary because the previous studies on spin-1/2 honeycomb magnets have focused mostly on the 4*d* and 5*d* transition metal systems such as $\alpha$-RuCl$_3$,[70,71] and the $A_2IrO_3$ iridates.[72,73] Similarly, it has been reported that GdInO$_3$ possesses no long-range magnetic ordering down to temperatures as low as 1.8 K, indicating magnetic frustrated characters at low temperatures.[64,74] However, more thorough studies with various experimental and theoretical methods need to be conducted to uncover the magnetic structure at the ground state.

In this article, the magnetic and MCE were studied in polycrystalline GdInO$_3$. The sharp upturn and the negative Curie-Weiss temperature indicate the onset of antiferromagnetic ordering below $T_N \sim 2.1$ K in GdInO$_3$. Negligible thermal and magnetic hysteresis indicate a reversible MCE. For the magnetic field changes of 0–50 kOe and 0–70 kOe, the maximum magnetic entropy change values are 9.65 J/kg K and 18.37 J/kg K, respectively, near the liquid helium temperature, with the corresponding relative cooling power (*RCP*) values of 115.01 J/kg and 211.31 J/kg. The large reversible MCE of polycrystalline GdInO$_3$ will provide another avenue for magnetocaloric refrigerant exploitation in frustrated magnetic systems near the cryogenic temperature region and will aid in understanding the exotic magnetic properties of the frustrated magnetic ground state in stuffed honeycomb antiferromagnet GdInO$_3$.

**Experiments**

The GdInO$_3$ powders were synthesized with the conventional solid-state reaction method. High purity Gd$_2$O$_3$ (99.99 %, General Research Institute for Nonferrous Metals) and In$_2$O$_3$ (99.99 %, GRINM) powders were thoroughly mixed in an equimolar amount and heated at 1673 K for 72 h with several intermediate grindings. A Rigaku Miniflex 624 X-ray diffractometer was used to carefully check the phase purities and crystal structures of the powders with high-intensity graphite monochromatized Cu-K$\alpha$ radiation at room temperature. The Rietveld refinement of the X-ray diffraction (XRD) pattern was carried out using Rietica software. The temperature and the magnetic field dependences of the magnetizations were measured for powder samles with masses of ~2.6 mg by a Quantum Design superconducting quantum interference device vibrating sample magnetometer (SQUID-VSM) within the temperature range of 1.8–300 K.

## Results and Discussion

The XRD measurement was carried out to confirm the phase purity of the synthesized powder. Fig. 2(c) displays the room-temperature powder XRD pattern of the synthesized sample with Rietveld refinement. The sample exhibits a hexagonal structure with a *P6₃cm* (No. 185) space group with refinement factors of $R_p = 4.728$ and $R_{wp} = 6.267$, and no impurity was detected. All the diffraction peaks in the XRD pattern coincide with the spectral lines of the GdInO₃ standard card (No. 14–0150) in the Joint Committee on Powder Diffraction Standards (JCPDS) database as shown in Fig. 2(d). The refined lattice parameters $a = 6.3569 \text{ Å}$ and $c = 12.3512 \text{ Å}$ agree well with the previously reported values.[74]

Figure 2 shows the temperature dependences of the magnetizations (the left vertical coordinates) in the magnetic fields of 0.1 kOe, 1 kOe, and 2 kOe. In the field of 0.1 kOe, the magnetization data were measured for three circumstances. The ZFC and FC curves represent the data recorded upon warming up after being zero-field cooled (ZFC) and field cooled (FC) from room temperature. In comparison, the FCC curve denotes the data upon cooling down from room temperature for a magnetic field of 0.1 kOe. However, for the magnetic fields of 1 kOe and 2 kOe, the temperature-dependent magnetization measurements were only performed under ZFC conditions. No splitting was observed between the ZFC, FC, and FCC curves for 0.1 kOe within the measured temperature range of 1.8–300 K, indicating a negligible thermal hysteresis in this material. The Curie-Weiss fitting of the *M*(*T*) data under 1 kOe was performed. This fitting is shown in the right vertical coordinates of Fig. 2. The deduced effective magnetic moment $\mu_{eff}$ 7.95 $\mu_B$ / Gd is consistent with the theoretical free Gd³⁺ ion effective moment ($\mu_{theo} = g_J\sqrt{J(J+1)}\mu_B = 7.94\mu_B$ with $g_J = 2$, $J = S = \frac{7}{2}$). The negative value of the extrapolated Curie-Weiss temperature $\theta_w$ (–7.41 K) indicates an antiferromagnetic feature of the magnetic Gd³⁺ moment correlations. A sharp upturn was observed in the *M*(*T*) curves for all the three measured magnetic fields (0.1, 1, and 2 kOe) at ~ 2.1 K, which was also verified by the peaks at ~2.1 K in the curves of the first derivative *M*(*T*) at 0.1 kOe, 1 kOe, and 2 kOe, as shown in the inset of Fig. 2. All the above phenomena suggest the onset of antiferromagnetic ordering below $T_N$ ~2.1 K in the GdInO$_3$.

The magnetic field-dependent magnetization hysteresis loops (*M*–*H*) and $\partial M/\partial H$ curves at 1.8 K and 10 K are shown in Fig. 3. The absence of hysteresis indicates a reversible MCE. With the increase of the magnetic field, the growth rate of the magnetizations at 1.8 K undergoes the first declining tendency in the magnetic field region of 0–28 kOe, followed by a rising tendency in the 28–46 kOe field region, and then a second declining tendency in the 46–70 kOe field region. Correspondingly, we observed the anomalies in the *M-H* hysteresis loop at ±28 kOe with the magnetization value of 2.081 $\mu_B/f.u.$ (~0.26 times the $\mu_{theo}$ value of the free Gd$^{3+}$ moment). This can be understood as the *M*$_s$/3 phenomenon induced by the collinear "up-up-down" arrangement of the Gd$^{3+}$ moments in the external magnetic field of the 2D triangular lattices. Similar results were also reported in other geometrically frustrated systems with 2D triangular magnetic lattices.[75-78] However, the growth rate of the magnetizations at 10 K has a declining tendency in the entire measured magnetic field range of 0–70 kOe, which demonstrates that the magnetic field induced the thermal fluctuation suppresses transition at high temperatures. At 1.8 K, the magnetization value reaches 4.81 $\mu_B/f.u.$ (107.47 emu/g) at 70 kOe, which corresponds to only ~60.5% of the expected theoretical Gd$^{3+}$ ion magnetic moments (7.94 $\mu_B/f.u.$). The isothermal magnetization curves in the temperature range of 2–60 K are shown in Fig. 4 (a). Compared to the other reported Gd$^{3+}$-based compounds such as GdScO$_3$,[44] GdFeO$_3$,[79] and GdCoO$_3$,[80] the field-dependent magnetizations of GdInO$_3$ reveal a slower growth rate and much lower values at the same magnetic fields. Even at a low temperature of 2 K and a high magnetic field of 70 kOe, no saturation value of the magnetization was observed. The above phenomena can be attributed to the stuffed honeycomb lattice-induced magnetic frustrations. Based on the isothermal magnetizations shown in Fig. 4(a), the Arrott plot of GdInO$_3$ was obatined, as shown in Fig. 4(b). All the data lines display positive slopes, which confirm a second-order characteristic of magnetic field induced magnetic transition according to the Banerjee criterion.[81]

Considering the prominent field-dependent magnetizations with negligible thermal and magnetization hysteresis, there can be an expectation of a large reversible MCE in polycrystalline GdInO$_3$. By using the Maxwell relationship, the magnetic entropy changes $-\Delta S_M$ were calculated based on the *M-H* data, as shown in the following equation:[5]

$$\Delta S_M(T,H) = \int_0^H (\frac{\partial S}{\partial H})_T dH = \int_0^H (\frac{\partial M}{\partial T})_H dH = \sum_0^H (\frac{M_{T_1} - M_{T_2}}{T_1 - T_2})\Delta H \quad , \quad (1)$$

where $M_{T_1}$ and $M_{T_2}$ represent the magnetizations at temperatures $T_1$ and $T_2$, respectively, for a magnetic field change of $\Delta H$. Fig. 5(a) shows the temperature dependences of the magnetic entropy changes in GdInO$_3$. The maximums of the magnetic entropy changes lie are maintained temperatures below 4 K, which then shift to higher temperatures with the increasing magnetic field change. Only the hot half parts of $-\Delta S_M$ peaks are observed, which may have been attributed to the highly geometrical frustrated magnetic lattices in GdInO$_3$. In the magnetic field changes of 0–20 kOe, 0–50 kOe, and 0–70 kOe, the maximum magnetic entropy values are 3.94 J/kg K at 2.5 K, 9.65 J/kg K at 3.5 K, and 18.37 J/kg K at 3.5 K, respectively. Additionally, the other two crucial parameters used to estimate the MCE performance are the refrigerant capacity (*RC*) and *RCP*. For convenience, these parameters are called *R* factors in this paper, and they are defined as follows:[82]

$$RC = -\int_{T_1}^{T_2} \Delta S_M(T) dT , \quad (2)$$

$$RCP = -\Delta S_M^{\max} \times \delta T_{FWHM} \quad (3)$$

where $-\Delta S_M^{\max}$, $T_1$, $T_2$, and $\delta T_{FWHM}$ represent the maximum value of $-\Delta S_M$, the cold end, the hot end, and the temperature interval mark at the half maximum $-\Delta S_M$, respectively. Because the cold temperature ends of the half $-\Delta S_M^{max}$ values are below the lowest temperature point (2.5 K) in the $S(T)$ curves, we adopted 2.5 K as $T_1$, and we obtained the $R$ factors for different magnetic fields, as shown in the inset of Fig. 5(b). For better understanding, the schematic diagrams of $RC$ and $RCP$ for the magnetic field change from 0 to 70 kOe are also illustrated in the inset of Fig. 5(a) and Fig. 5(b). For the field changes of 0–50 kOe and 0–70 kOe, the $RC$ values are 89.06 J/kg and 161.29 J/kg, respectively, and the $RCP$ values are 115.01 J/kg and 211.31 J/kg. For comparison, the transition temperatures and the MCE parameters of GdInO$_3$ and other cryogenic magnetocaloric materials were recorded, as shown in Table 1.[42,44-49,52,79,80,83-94] GdInO$_3$ demonstrates a much lower magnetic entropy change value than the theoretical limiting value of 54.01 J/kg K (deduced from $\Delta S_{max} = R\ln(2J + 1)$, where $R$ and $J$ represent the gas constant and the total angular momentum of the magnetic ion, respectively) as well as other Gd$^{3+}$-based or Eu$^{2+}$-based compounds such as GdScO$_3$,[44,95] GdCrO$_3$,[35] GdAlO$_3$,[46] EuTiO$_3$,[51,52] GdVO$_4$,[47,48] and GdPO$_4$,[49] which may have resulted from the stuffed honeycomb lattice induced magnetic frustrations between Gd$^{3+}$ moments. However, the maximum magnetic entropy change $-\Delta S_M^{max}$ (18.65 J/kg K) and the $R$ factors (161.29 J/kg for $RC$, 211.31 J/kg for $RCP$) with the magnetic field change of 0–70 kOe still show a comparable or even larger value than some other rare-earth-based compounds near the liquid helium temperature, such as $R_6$Co$_2$Ga ($R$ = Gd, Ho),[96] $R_2$Mo$_2$O$_7$ ($R$ = Gd, Er, Dy),[42] DyVWO$_6$,[91] $R_2$CrMnO$_6$ ($R$ = Ho and Er),[93] $R_2$FeCrO$_6$ ($R$ = Er and Tm),[97] $R$VO$_4$ ($R$ = Er, Ho and Yb),[47] and $R$AgAl ($R$ = Er, Ho).[92] The experimental results in our work indicate that GdInO$_3$ possesses a reversible MCE near the liquid helium temperature. This offers great potential for the applications of frustrated magnetic systems in future low-temperature magnetic refrigeration technology, and will aid the exploration of the complicated mechanisms in frustrated stuffed honeycomb magnetic lattices.

## Conclusions

In summary, we performed magnetic and magnetocaloric investigations of polycrystalline GdInO$_3$. The onset of antiferromagnetic ordering was observed at $T_N$ ~ 2.1 K. The magnetization value of only 60.5% of free Gd$^{3+}$ moments at 1.8 K and 70 kOe reveals frustration in this stuffed honeycomb antiferromagnet system. No thermal and magnetization hysteresis was observed, which indicates a reversible MCE performance. In the magnetic field changes of 0–50 kOe and 0–70 kOe, the maximum magnetic entropy changes of 9.65 J/kg K and 18.37 J/kg K, respectively, were obtained near the liquid helium temperature. Correspondingly, the *RC* values are 89.06 J/kg and 161.29 J/kg, and the *RCP* values are 115.01 J/kg and 211.31 J/kg. The relatively lower maximum magnetic entropy changes of GdInO$_3$ compared to the theoretical changes may have been because the triangular-honeycomb magnetic Gd$^{3+}$ lattice caused magnetic frustration. The MCE performance is expected to facilitate understanding of the frustrated magnetic interactions in stuffed honeycomb antiferromagnet GdInO$_3$.

## Author contributions

Y.-D. W. supervised the project, conceived the idea, and designed the experiment. Y.-D. W., W.-W. D., Q.-Y. L. synthesized the GdInO$_3$ polycrystalline. Y.-D. W., W.-W. D., Q.-Y. L. made the XRD measurement and performed the Rietveld refinement. Y.-D. W. and Y.-L. Q. made the magnetic measurements. Y.-D. W. carried out the collection, processed, and analysis of data for the work. Y.-D. W., Z.-F. Z., and J. T. wrote the paper with input from all the co-authors. All authors discussed the results and commented on the manuscript.

Acknowledgments

This work was supported by the National Natural Sciences Foundation of China (Grant No. 12104123, U1632161), Anhui Provincial Funds for Distinguished Young Scientists of the Nature Science (Grant No. 1808085JQ13), the Natural Science Foundation of Anhui Province (Grant No. 2008085MF217), Universities Joint Key Laboratory of Photoelectric Detection Science and Technology in Anhui Province (Grant No. 2019GDTC06), the open fund project from Anhui Province Key Laboratory of simulation and design for Electronic information system (Grant No. 2019ZDSYSZY04), the Project of Leading Backbone Talents in Anhui Provincial Undergraduate Universities, and Undergraduate Innovation and Entrepreneurship Training Program in Anhui Province (Grant No. S202014098164).

## Table

**Table 1** The phase transition temperature ($T_M$) and the maximum magnetic entropy changes $-\Delta S_M^{max}$ of rare-earth-based oxides with the magnetic field changes of 0–20 kOe, 0–50 kOe, and 0–70 kOe. SG and PO represent single crystals and polycrystals, respectively. The "--" symbols illustrate that the related results have not been reported.

| Material | $T_M$ (K) | $-\Delta S_M^{max}$ (J/kg K) | | | R factors (0-70 kOe) (J/kg) | | Reference |
|---|---|---|---|---|---|---|---|
| | | 0–20 kOe | 0–50 kOe | 0–70 kOe | RC | RCP | |
| **GdInO$_3$** | **2.1** | **3.94** | **9.65** | **18.37** | **161.29** | **211.31** | **this work** |
| GdScO$_3$ | 2.8 | 7.90 | 39.39 | 50.77 | 325.11 | 436.61 | [44] |
| GdFeO$_3$ | 3.6 | ~8 | ~38 | 50.2 | 321.2 | -- | [79] |
| GdMnO$_3$ | 4.5 | ~5 | ~12 | ~18 | -- | 211 | [46] |
| GdAlO$_3$ | 3.9 | ~4 | ~23 | ~35 | -- | 203 | [46] |
| GdCrO$_3$ | 2.3 | -- | ~30 | 36.97 | -- | 542 | [45] |
| GdCrO$_4$ | 21.3 | -- | 18.8 | -- | -- | -- | [83] |
| EuTiO$_3$ | 5.6 | 21.1 | 40.4 | 49 | 500 | -- | [52,84] |
| GdPO$_4$ | 0.77 | -- | -- | 62.0 | -- | -- | [49] |
| GdVO$_4$ | 2.50 | | 41.1 | 63 | -- | -- | [48] |
| GdCoO$_3$ | 3.1 | ~5 | 26.7 | 39.1 | 278 | -- | [80] |
| Gd(OH)$_3$ | 27 | 26.9 | ~56 | 62.00 | -- | -- | [85] |
| Gd$_2$O(OH)$_4$(H$_2$O)$_2$ | 31 | 17.0 | ~52 | 59.09 | -- | -- | [85] |
| Gd$_2$Mo$_2$O$_7$ | 78 | 4.1 | 11.1 | 15.2 | 467.9 | -- | [42] |
| Gd$_2$ZnMnO$_6$ | 6.4 | 2.03 | 15.17 | 25.20 | 313 | 407 | [86] |
| Gd$_2$FeAlO$_6$ | < 2 | 5.2 | -- | 25.9 | 240.1 | -- | [87] |
| Sr$_2$GdNbO$_6$ | 2.0 | 16.35 | 26.07 | 29.71 | 188.1 | 267.4 | [88] |
| Gd$_6$Co$_2$Ga | 55/78/129 | ~3 | ~9 | 12.6 | -- | 945.3 | [89] |
| Ho$_6$Co$_2$Ga | 16 | ~2 | ~10 | 15.8 | -- | 532.6 | [89] |
| Ho$_2$TiMnO$_7$ | 7.5 | 5.98 | 11.45 | 13.95 | 573.2 | -- | [90] |
| HoVO$_4$ | 3.5 | ~2 | 7.94 | -- | -- | -- | [47] |
| DyVWO$_6$ | 2.5/5 | ~3 | ~7 | 8.45 | -- | -- | [91] |
| ErAgAl | 14 | 4.2 | 10.5 | 13.7 | -- | 398 | [92] |
| HoAgAl | 18 | 3.8 | 10.3 | 13.8 | -- | 525 | [92] |
| Ho$_2$CrMnO$_6$ | 6.1 | 3.48 | ~9 | 12.94 | 246.4 | 322.7 | [93] |
| HoTiO$_3$ | 53 | 5.96 | 11.56 | -- | -- | -- | [94] |

# Figure captions

**Fig. 1.** (a) The hexagonal $P6_3cm$ crystal structure of $GdInO_3$. (b) The stuffed honeycomb arrangement of Gd1 and Gd2 sites in the *ab* plane. The green, blue, white, and red balls denote Gd1, Gd2, In, and O atoms, respectively. The thick, solid-colored sticks represent the atomic bonds. The solid black lines represent the crystal unit cells of $GdInO_3$. (c) XRD pattern of polycrystalline $GdInO_3$ measured at room temperature. The cross-shaped symbols and the solid red line represent the experimental data and the Rietveld refined results, respectively. The blue line represents the difference between the experiment and the fitting results. (d) Standard card (No. 14–0150) for powdered $GdInO_3$ in the Joint Committee on Powder Diffraction Standards (JCPDS) database.

**Fig. 2.** Left vertical coordinates: The temperature dependences of ZFC, FCC, and FC magnetizations in magnetic fields of 0.1 kOe, 1 kOe, and 2 kOe. Right vertical coordinates: The Curie-Weiss fitting of 1 kOe ZFC data in the temperature range of 50–300 K. The inset displays the temperature dependence of the first derivative values of the ZFC magnetizations (blue triangle symbols) of polycrystalline $GdInO_3$ in fields of 0.1 kOe, 1 kOe, and 2 kOe.

**Fig. 3.** (a) The magnetization hysteresis (*M-H*) loops measured at 1.8 K (blue line) and 10 K (red line) for polycrystalline $GdInO_3$ in a magnetic field range from –70 kOe to +70 kOe. (b) The first derivative curves of the magnetic field-dependent *M-H* loops at 1.8 K (blue square symbols) and 10 K (red circle symbols).

**Fig. 4.** (a) Magnetic field dependence of isothermal magnetizations in the temperature range of 2–60 K. (b) Arrott plot of polycrystalline $GdInO_3$ in the temperature range of 2–52 K.

**Fig. 5.** (a) Temperature dependence of magnetic entropy changes $-\Delta S_M^R$ of polycrystalline $GdInO_3$. (b) The magnetic field dependences of *RC* (magenta hexagon symbols) and *RCP* (blue star symbols) curves. The insets in Fig. 5(a) and Fig. 5(b) illustrate the schematic diagrams of *RCP* (the magenta area) and *RC* (the cyan area) for the magnetic field change of 0-70 kOe.

Figures:

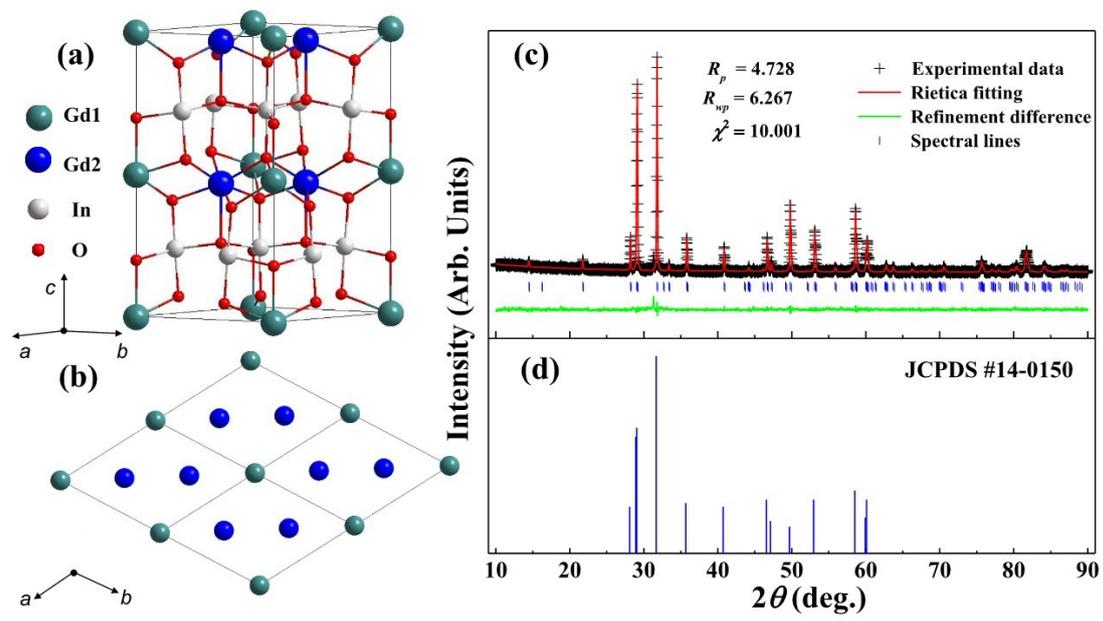

Fig. 1.

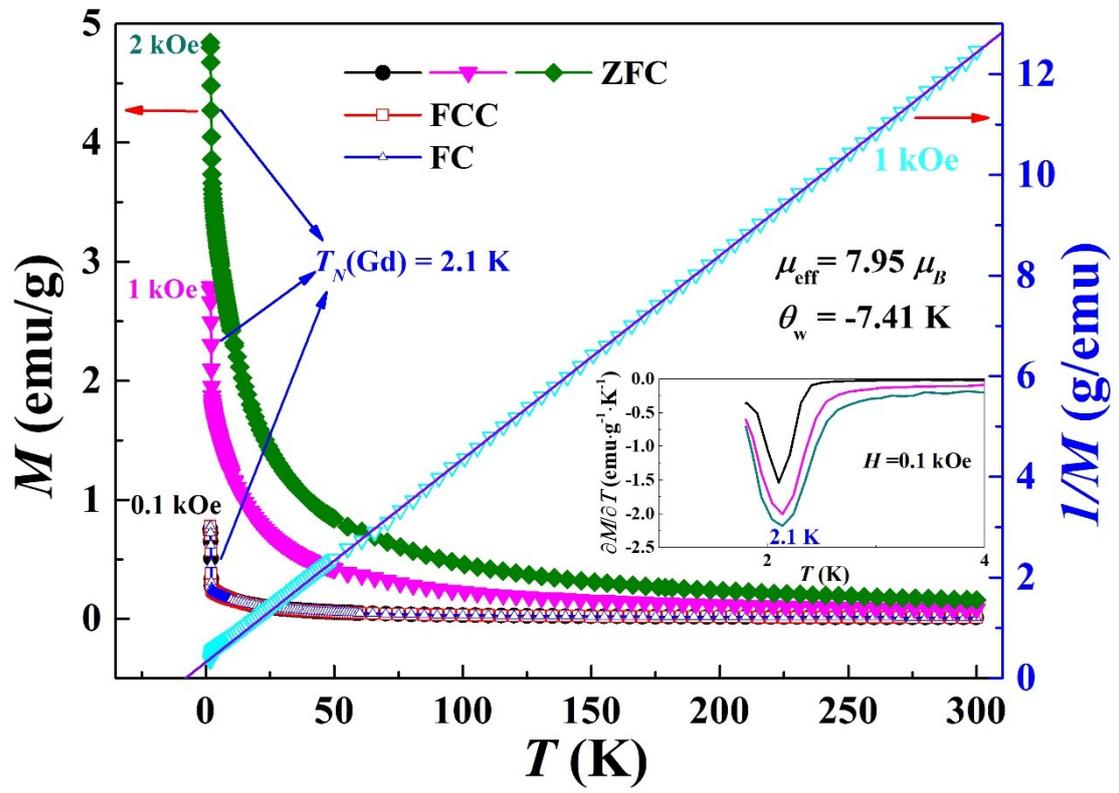

Fig. 2.

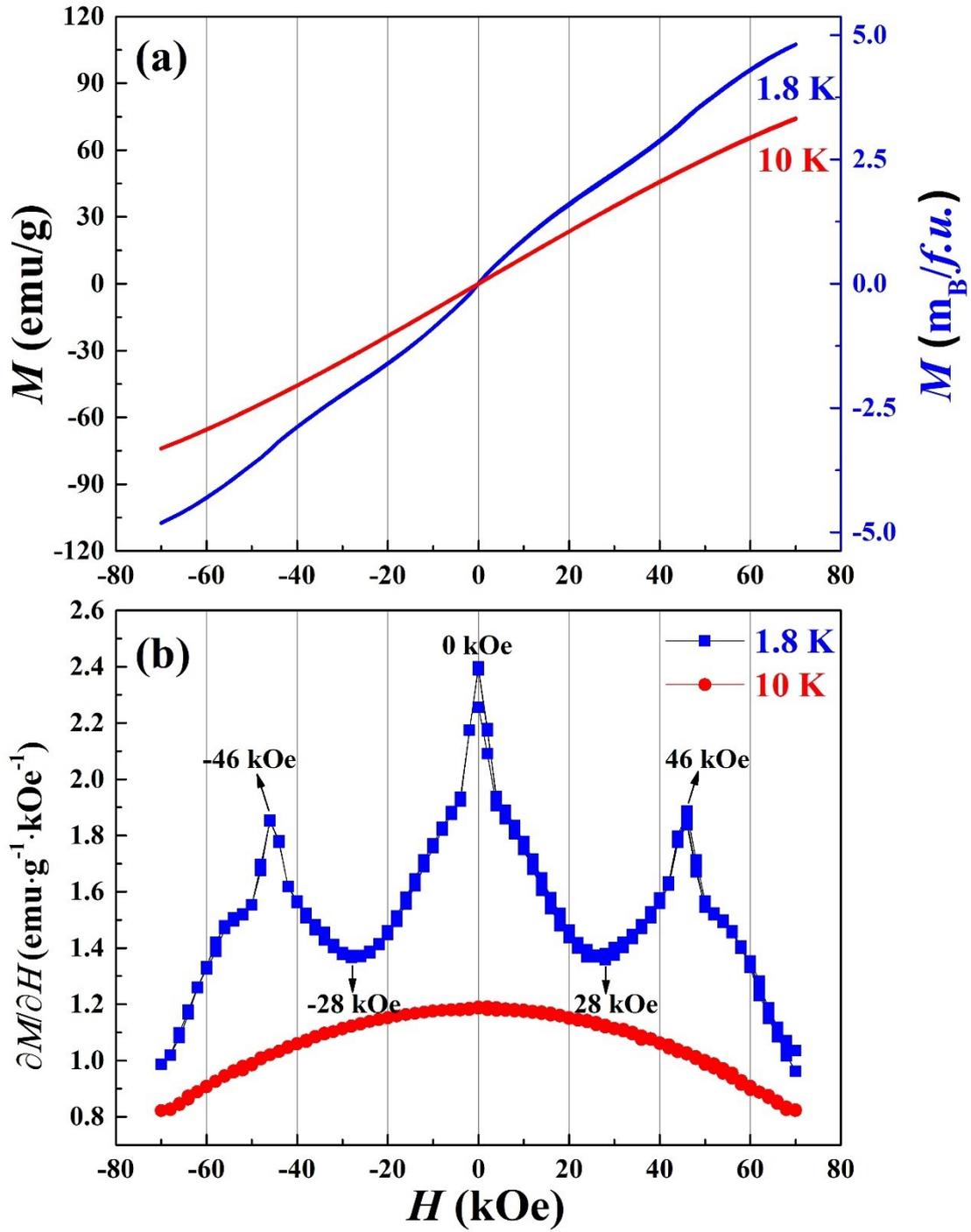

Fig. 3.

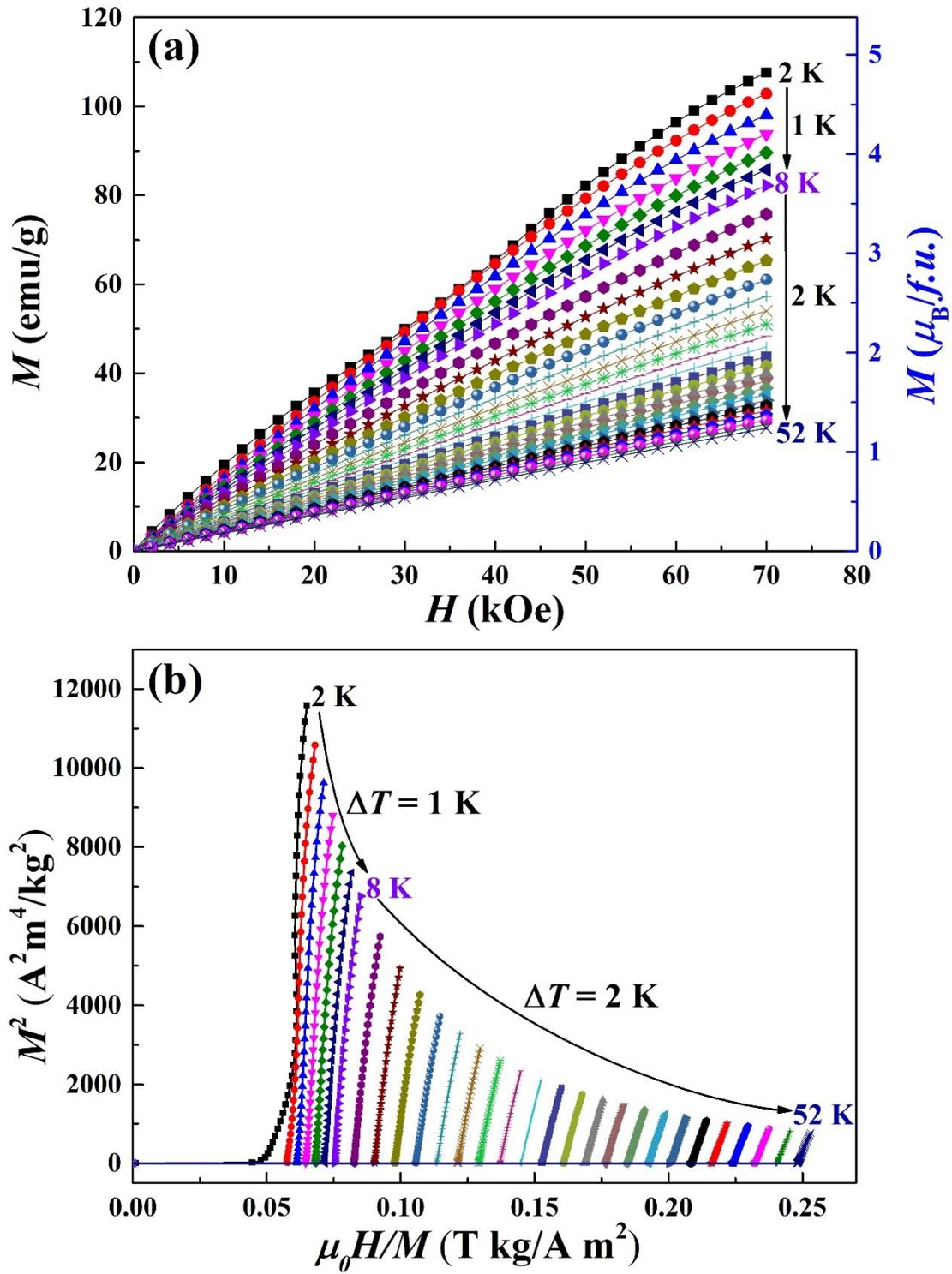

Fig. 4.

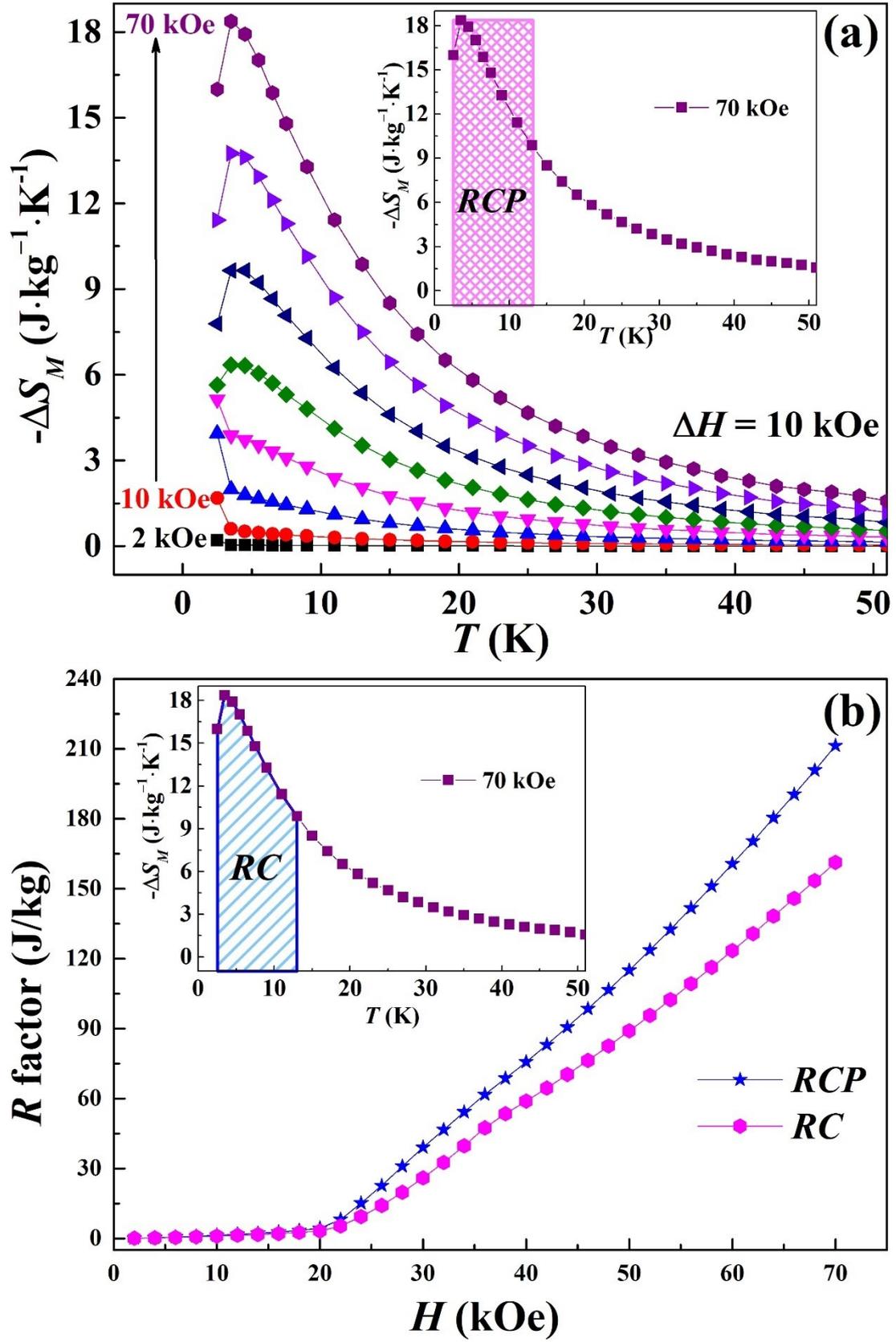

Fig. 5.

23